# New copulas based on general partitions-of-unity (part III) – the continuous case (extended version)

Dietmar Pfeifer[1], Andreas Mändle[2], Olena Ragulina[3] and Côme Girschig[4]

May 20, 2019

**Abstract** In this paper we discuss a natural extension of infinite discrete partition-of-unity copulas which were recently introduced in the literature to continuous partition of copulas with possible applications in risk management and other fields. We present a general simple algorithm to generate such copulas on the basis of the empirical copula from high-dimensional data sets. In particular, our constructions also allow for an implementation of positive tail dependence which sometimes is a desirable property of copula modelling, in particular for internal models under Solvency II.

**Keywords** Partition-of-unity copulas, tail dependence

## 1 Introduction

General discrete infinite partition-of-unity copulas have been introduced recently in a series of papers by Pfeifer et al. ([6], [7]). Such copula constructions comprise, in particular, Bernstein copulas and allow, among other advantages, for simple Monte Carlo studies in risk management on the basis of observed data without specific fitting procedures to parametric copula models. In particular, various kinds of tail dependence can be implemented into the copula construction if this seems to be appropriate for estimates of the risk measure for a portfolio of risks under consideration. The present paper completes our previous approaches by considering continuous infinite partition-of-unity copulas which have not yet been investigated before. We conclude the paper with a new study of a high dimensional data set from the insurance branch where we compare classical and partition-of-unity copulas with and without tail dependence.

## 2 Continuous partition-of-unity copulas

Assume that $\varphi_k(s,u)$ for $k=1,\cdots,d \in \mathbb{N}$ represent Lebesgue densities of distributions over $\mathbb{R}$ with a parameter $u \in (0,1)$, i.e.

(1)    $\varphi_k(s,u) \geq 0$   and   $\int\limits_{-\infty}^{\infty} \varphi_k(s,u)\,ds = 1$ for $u \in (0,1),$

and let

---
[1] Carl von Ossietzky Universität Oldenburg, Germany
[2] University of Bremen
[3] Taras Shevchenko National University of Kyiv, Ukraine
[4] École des Ponts ParisTech

(2) $\quad \alpha_k(s) := \int_0^1 \varphi_k(s,u)\, du \in (0,\infty), \ A_k(s) := \int_{-\infty}^s \alpha_k(w)\, dw \ \text{ for } s \in \mathbb{R}.$

Then by normalizing the $\varphi_k(s,u)$ w.r.t. $u \in (0,1)$, we obtain new densities

(3) $\quad f_k(s,u) := \dfrac{\varphi_k(s,u)}{\alpha_k(s)}, \ u \in (0,1) \ \text{ for } s \in \mathbb{R}.$

**Theorem 1.** Let $p(s_1, \cdots, s_d)$ denote the density of an arbitrary absolutely continuous $d$-dimensional random vector $\mathbf{S} = (S_1, \cdots, S_d)$ over $\mathbb{R}^d$ with marginal densities $\alpha_k(\bullet)$ for $S_k$. Then

(4) $\quad c(\mathbf{u}) := \int_{-\infty}^{\infty} \cdots \int_{-\infty}^{\infty} p(s_1, \cdots, s_d) \prod_{k=1}^{d} f_k(s_k, u_k)\, ds_1 \cdots ds_d, \ \mathbf{u} = (u_1, \cdots, u_d) \in (0,1)^d$

defines the density of a $d$-variate copula, which is called *continuous partition-of-unity copula* (CPU-copula for short).

The proof of Theorem 1 is completely analogous to the proofs of the corresponding theorems in Pfeifer et al. ([6], [7]) and is therefore omitted.

Note that relation (4) also remains valid if the distribution induced by $p$ is singular, i.e. if $P$ is a probability measure with marginal distributions that possess the densities $\alpha_1, \cdots, \alpha_d$, without having a density w.r.t. the $d$-dimensional Lebesgue measure. In this case, $c(\mathbf{u})$ is a singular mixture of product densities given by

(5) $\quad c(\mathbf{u}) := \int \prod_{k=1}^{d} f_k(\bullet, u_k)\, dP, \ \mathbf{u} = (u_1, \cdots, u_d) \in (0,1)^d.$

For instance, if $P$ corresponds to the upper Fréchet bound and all $\varphi_k$ are identical – hence also all $f_k$ are identical, say to $f$ – we have

(6) $\quad c(\mathbf{u}) := \int_{-\infty}^{\infty} \alpha(s) \prod_{k=1}^{d} f(s, u_k)\, ds, \ \mathbf{u} = (u_1, \cdots, u_d) \in (0,1)^d.$

We call this situation the case of *diagonal dominance*.

Note that it is extremely easy to perform a Monte Carlo simulation for a CPU-copula. We describe the procedure in the following steps.

**Step 1:** Let $(U_1, \cdots, U_d)$ be a vector of random variables with a given copula $\tilde{C}$ as joint distribution function, which we call *copula driver* for our construction. Let $Q_k(u) := A_k^{-1}(u), \ 0 < u < 1, \ k = 1, \cdots, d$ denote the quantile function pertaining to $\alpha_k$. Define $S_k := Q_k(U_k)$ for $k = 1, \cdots, d$. Then $(S_1, \cdots, S_d)$ possesses the joint (possibly degenerated) distribution $P$ with the desired marginal distributions (in fact, $\tilde{C}$ here is the copula pertaining to $P$).

**Step 2:** Let $(s_1, \cdots, s_d)$ be a realization of $(S_1, \cdots, S_d)$ according to Step 1. Let further $v_1, \cdots, v_d$ be independent realizations of the distributions with marginal densities $f_k(s_k, \bullet)$ for $k = 1, \cdots, d$. Then $(v_1, \cdots, v_d)$ is a realization of a random vector $(V_1, \cdots, V_d)$ whose distribution is given by the CPU-copula with the density $c$ as given by (4) or (5).

A particularly interesting choice of a data-driven $\tilde{C}$ is a copula that is derived from the empirical copula in the sense of Deheuvels [2]. Such an approach was discussed in Pfeifer et. al. [7]. In particular, the following type of a driver is of importance, which is constructed as a patchwork copula with a local Gaussian copula. Suppose that $n$ independent observations $(\mathbf{z}_1, \cdots, \mathbf{z}_n)$ of a multivariate random vector $\mathbf{Z} = (Z_1, \cdots, Z_d)$ with an absolutely continuous distribution are given. Without loss of generality, we can identify the pertaining empirical copula with the empirical rank vectors $(\mathbf{r}_1, \cdots, \mathbf{r}_n)$ where $r_{ik}$, $i = 1, \cdots, n$, $k = 1, \cdots, d$ is the rank of $z_{ik}$ among $z_{1k}, \cdots, z_{nk}$. Let $C_i^G$ for $i = 1, \cdots, n$ denote a $d$-dimensional Gaussian copula with variance-covariance matrix $\Sigma_i$, and $Y_1, \cdots, Y_n$ be independent random vectors with joint distribution $C_i^G$, $i = 1, \cdots, n$. Let further $J$ denote a random variable which is uniformly distributed over the set $\{1, \cdots, n\}$, independent of the $Y_1, \cdots, Y_n$. Then the random vector $\mathbf{W} := \frac{1}{n}(\mathbf{Y}_J + \mathbf{r}_J - \mathbf{1})$ with $\mathbf{1} = (1, \cdots, 1)$ ($d$ times) possesses a patchwork copula $\tilde{C}$ as joint distribution, which is concentrated around the relative ranks of the data. $\tilde{C}$ can be considered as a kind of natural extension of the empirical copula to a true copula, which is close to the original dependence structure of the data. For the choice of $\Sigma_i = diag(1, \cdots, 1)$, which corresponds to the independent case, we obtain a rook copula (cf. Cottin and Pfeifer [1]). For the choice of $\Sigma_i = \begin{pmatrix} 1 & \cdots & 1 \\ \vdots & \ddots & \vdots \\ 1 & \cdots & 1 \end{pmatrix}$, we obtain the upper Fréchet bound as a driver. In two dimensions, $\Sigma_i = \begin{pmatrix} 1 & -1 \\ -1 & 1 \end{pmatrix}$ gives the lower Fréchet bound as a driver.

We call the resulting copula drivers UF and LF copula drivers, resp. The independent case (i.e. with zero correlations) is called rook copula driver.

## 3 Particular cases

Firstly, we introduce what we call the *Gamma copula* model. To start with, denote $L(u) := -\ln(u)$, $0 < u < 1$, and define $\varphi_k(s, u) := \frac{L^{a_k}(u)}{\Gamma(a_k)} s^{a_k - 1} u^s$ $0 < u < 1, s > 0, k = 1, \cdots, d$ with given parameters $a_1, \cdots, a_d > 0$. Hence the $\varphi_k(\bullet, u)$ are densities of Gamma distributions with parameters $a_k$ and $1/L(u)$ with the notation of Klugman et al. [3], A.3.2.1. Substituting $x = L(u)$ or $u = e^{-x}$, i.e. $du = -e^{-x} dx$, we get

$$(7) \quad \alpha_k(s) := \int_0^1 \varphi_k(s,u)\,du = \int_0^\infty \frac{x^{a_k}}{\Gamma(a_k)} s^{a_k-1} e^{-(1+s)x}\,dx$$

$$= a_k \frac{s^{a_k-1}}{(1+s)^{a_k+1}} \int_0^\infty \frac{x^{a_k}}{\Gamma(a_k+1)} (1+s)^{a_k+1} e^{-(1+s)x}\,dx = a_k \frac{s^{a_k-1}}{(1+s)^{a_k+1}}, \quad s > 0.$$

Note that the $\alpha_k$ are the densities of inverse Pareto distributions with parameters $a_k$ and 1 (in the notation of Klugman et al. [3], A.2.3.2). With

$$(8) \quad f_k(s,u) := \frac{\varphi_k(s,u)}{\alpha_k(s)} = \frac{(1+s)^{a_k+1}}{\Gamma(a_k+1)} L^{a_k}(u) u^s$$

we obtain from (4) the density of a *Gamma copula*:

$$(9) \quad c_{\mathbf{a}}^{\Gamma}(\mathbf{u}) := \int_0^\infty \cdots \int_0^\infty p(s_1,\cdots,s_d) \prod_{k=1}^d f_k(s_k, u_k)\,ds_1 \cdots ds_d, \quad 0 < u_1, \cdots, u_d < 1$$

with the $f_k$ given in (8). Here $p$ is the density of an absolutely continuous multivariate distribution with marginal densities $\alpha_k$, $k = 1,\cdots,d$, and $\mathbf{a} := (a_1,\cdots,a_d)$ denotes the vector of parameters. A corresponding modification for the singular case discussed above is obvious.

Note that in the singular case of two-dimensional diagonal dominance, we get, with $\mathbf{a} = (a,a)$,

$$(10) \quad c_{\mathbf{a},\mathrm{sing}}^{\Gamma}(u,v) := \int_0^\infty \alpha(s) f(s,u) f(s,v)\,ds = \int_0^\infty \frac{a\,s^{a-1}}{(1+s)^{a+1}} \left[\frac{(1+s)^{a+1}}{\Gamma(a+1)}\right]^2 L^a(u) u^s L^a(v) v^s\,ds$$

$$= \frac{\left((-\ln(u))(-\ln(v))\right)^a}{\Gamma(a)\Gamma(a+1)} \int_0^\infty s^{a-1}(1+s)^{a+1}(uv)^s\,ds, \quad 0 < u, v < 1.$$

For integer values of $a$, this can be simplified to

$$(11) \quad c_{\mathbf{a},\mathrm{sing}}^{\Gamma}(u,v) = \frac{\{\ln(u)\ln(v)\}^a}{\{-\ln(uv)\}^{2a+1}} \cdot \sum_{i=0}^{a+1} \binom{a+1}{i} \frac{(a+i-1)!}{a!(a-1)!} \{-\ln(uv)\}^{a+1-i}, \quad 0 < u, v < 1.$$

*Proof.* We have $\displaystyle\int_0^\infty s^{a-1}(1+s)^{a+1}(uv)^s\,ds = \sum_{k=0}^{a+1} \binom{a+1}{k} \int_0^\infty s^{a-1} s^k (uv)^s\,ds = \sum_{k=0}^{a+1} \binom{a+1}{k} \int_0^\infty s^{a-1+k}(uv)^s\,ds.$

The final result follows from the observation

$$\int_0^\infty s^m z^s\,ds = \int_0^\infty s^m e^{-ys}\,ds = \frac{1}{y^{m+1}} \int_0^\infty y^{m+1} s^m e^{-ys}\,ds = \frac{1}{y^{m+1}} \int_0^\infty t^m e^{-t}\,dt = \frac{\Gamma(m+1)}{(-\ln z)^{m+1}}$$

for every $m > 0$, $z \in (0,1)$ with the substitution $y = -\ln z$. ∎

Note also that the $f_k(s,\bullet) = \dfrac{\varphi_k(s,\bullet)}{\alpha_k(s)}$ are densities of exponentially transformed Gamma distributions. To be more precise, let $X_k$ be a Gamma distributed random variable with parameter $a_k+1$ and $1/(s_k+1)$. The corresponding density is thus given by

$$(12) \quad f_{X_k}(x) = \frac{(1+s_k)^{a_k+1}}{\Gamma(a_k+1)} x^{a_k} e^{-(1+s_k)x}, \quad x>0,$$

and the random variable $U_k := e^{-X_k}$ possesses the density $f_k(s_k,\bullet)$. This follows from the observation that $F_{U_k}(u) := P(U_k \leq u) = P(X_k \geq -\ln(u)) = 1 - F_{X_k}(-\ln(u))$, $0<u<1$, hence

$$(13) \quad f_{U_k}(u) := \frac{d}{du} F_{U_k}(u) = \frac{d}{du}\left(1 - F_{X_k}(-\ln(u))\right) = \frac{f_{X_k}(-\ln(u))}{u}$$

$$= \frac{(1+s_k)^{a_k+1}}{\Gamma(a_k+1)} L^{a_k}(u) u^{s_k} = f(s_k,u), \quad 0<u<1,$$

A Monte Carlo simulation for a random vector $(V_1,\cdots,V_d)$ with a general Gamma copula density $c_{\mathbf{a}}^{\Gamma}$ (singular or not) is straightforward since here $Q_{a_k}(u) := \dfrac{u^{1/a_k}}{1-u^{1/a_k}}$, $0<u<1$, $k=1,\cdots,d$.

Secondly, we introduce what we call the *Power copula* model. Here we define

$$(14) \quad \varphi_k(s,u) := \begin{cases} \beta_k \left(\dfrac{s}{u}\right)^{\beta_k-1}, & s \leq u, \\ \beta_k \left(\dfrac{1-s}{1-u}\right)^{\beta_k-1}, & s>u, \\ 0, & \text{otherwise,} \end{cases} \quad 0<u,s<1, \ k=1,\cdots,d$$

with given parameters $\beta_1,\cdots,\beta_d > 2$. It is straightforward to see that

$$(15) \quad \alpha_k(s) = \int_0^1 \varphi_k(s,u)\,du = \frac{\beta_k}{\beta_k-2}\left[1 - s^{\beta_k-1} - (1-s)^{\beta_k-1}\right], \quad 0 \leq s \leq 1$$

and

$$(16) \quad A_k(s) = \int_0^s \alpha_k(t)\,dt = \frac{(1-s)^{\beta_k} - s^{\beta_k} + \beta_k s - 1}{\beta_k - 2}, \quad 0 \leq s \leq 1.$$

With

$$
(17) \quad f_k(s,u) := \frac{\varphi_k(s,u)}{\alpha_k(s)} = \begin{cases} \dfrac{(\beta_k - 2)\left(\dfrac{1-s}{1-u}\right)^{\beta_k - 1}}{1 - s^{\beta_k - 1} - (1-s)^{\beta_k - 1}}, & u \leq s, \\[2ex] \dfrac{(\beta_k - 2)\left(\dfrac{s}{u}\right)^{\beta_k - 1}}{1 - s^{\beta_k - 1} - (1-s)^{\beta_k - 1}}, & u > s, \\[2ex] 0, & \text{otherwise,} \end{cases}
$$

we obtain the density of a *Power copula*:

$$
(18) \quad c_{\boldsymbol{\beta}}^P(\mathbf{u}) := \int_0^1 \cdots \int_0^1 p(s_1, \cdots, s_d) \prod_{k=1}^d f_k(s_k, u_k) \, ds_1 \cdots ds_d, \; 0 < u_1, \cdots, u_d < 1
$$

where the $f_k$ are given in (17). Here $p$ again is the density of an absolutely continuous multivariate distribution with marginal densities $\alpha_k$, $k = 1, \cdots, d$. A corresponding modification for the singular case discussed above is obvious.

The following graphs show the densities $f_k(s,u)$ for different values of $\beta$, for $s = \dfrac{k}{10}$, $k = 1 \ldots 9$.

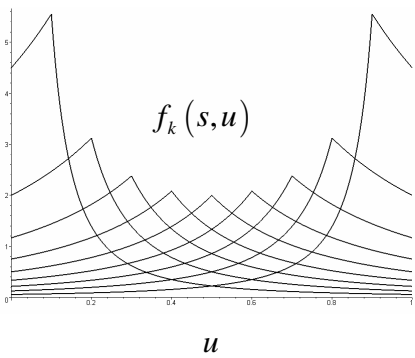
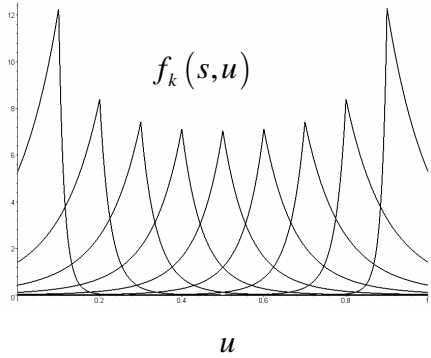
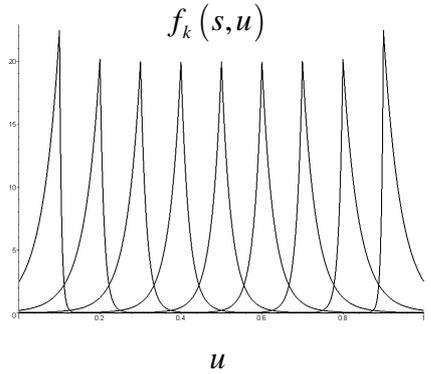

Fig. 1: $\beta = 3$           Fig.2: $\beta = 9$           Fig.3: $\beta = 22$

Note that in the case of two-dimensional diagonal dominance, we do not obtain a closed form representation of the corresponding copula density. However, a Monte Carlo simulation of the Power copula in arbitrary dimensions is easy since an elementary integration shows that there holds

$$
(19) \quad F_k(s,u) := \int_0^s f_k(t,u) \, dt = \begin{cases} 0, & s < 0, \\[1ex] \dfrac{(1-s)^{\beta_k - 1}\left[(1-u)^{2-\beta_k} - 1\right]}{1 - s^{\beta_k - 1} - (1-s)^{\beta_k - 1}}, & u \leq s, \\[2ex] \dfrac{1 - (1-s)^{\beta_k - 1} - s^{\beta_k - 1} u^{2-\beta_k}}{1 - s^{\beta_k - 1} - (1-s)^{\beta_k - 1}}, & s < u \leq 1, \\[2ex] 1, & s > 1, \end{cases}
$$

with the corresponding quantile functions $Q_k(s, \bullet) = F_k^{-1}(s, \bullet)$ given by

$$(20) \quad Q_k(s,u) := \begin{cases} \left[1 - \left[\dfrac{(1-s)^{\beta_k-1}}{(1-s)^{\beta_k-1} + u\left(1 - s^{\beta_k-1} - (1-s)^{\beta_k-1}\right)}\right]^{1/(\beta_k-2)}\right], & u \leq F_k(s,\beta_k,s), \\ \left[\dfrac{s^{\beta_k-1}}{1 - (1-s)^{\beta_k-1} - u\left(1 - s^{\beta_k-1} - (1-s)^{\beta_k-1}\right)}\right]^{1/(\beta_k-2)}, & u > F_k(s,\beta_k,s), \end{cases} \quad 0 < u < 1,$$

where

$$(21) \quad F_k(s,\beta_k,s) = \frac{1 - (1-s)^{\beta_k-1} - s}{1 - s^{\beta_k-1} - (1-s)^{\beta_k-1}}, \quad 0 \leq s \leq 1.$$

For the simulation of a random variable following the distribution with cdf $A_k$, a simple tabular inversion method is appropriate.

## 4 Tail dependence

The Gamma copula shows an upper tail dependence that coincides precisely with that of the negative binomial copula $c^{NB}$, a particular discrete partition-of-unity copula, see Pfeifer et al. ([6],[7]):

$$(22) \quad c_{\mathbf{a},\text{sing}}^{NB}(u,v) := (a+1)\frac{\left((1-u)(1-v)\right)^a}{(1-uv)^{2a+1}} \sum_{i=0}^{a-1} \binom{a-1}{i}\binom{a+1}{i}(uv)^i, \quad u,v \in (0,1), \ a \in \mathbb{N}.$$

Fig. 4 to Fig. 7 show the ratios of negative binomial and Gamma copula densities, for various values of $a$, which also suggest that these copula types are tail-equivalent. In fact, since $\dfrac{-\ln(1-h)}{h} = 1 + \mathcal{O}(h)$ for $h \downarrow 0$, we see by a comparison of (11) and (22) that

$$(23) \quad \lim_{\substack{u \to 1 \\ v \to 1}} \frac{c_{\mathbf{a},\text{sing}}^{NB}(u,v)}{c_{\mathbf{a},\text{sing}}^{\Gamma}(u,v)} = \frac{(a+1)\sum_{i=0}^{a-1}\binom{a-1}{i}\binom{a+1}{i}}{\dfrac{(2a)!}{a!(a-1)!}} = \frac{\sum_{i=0}^{a-1}\binom{a+1}{i}\binom{a-1}{a-1-i}}{\binom{2a}{a-1}} = 1$$

(cf. Stanley [8], Example 1.1.17, p. 12).

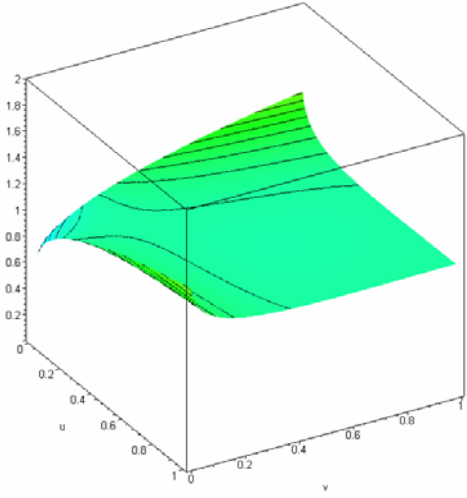 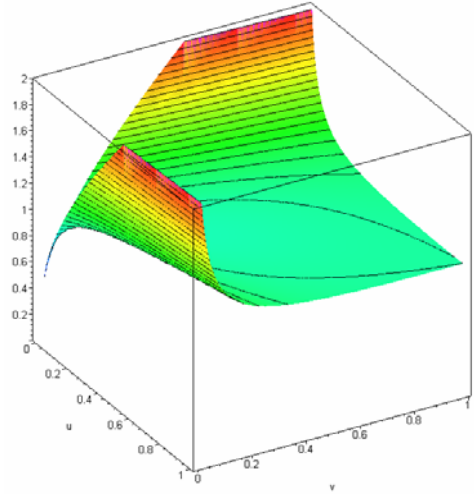

Fig. 4: $\dfrac{c^{NB}_{\mathbf{a},\text{sing}}(u,v)}{c^{\Gamma}_{\mathbf{a},\text{sing}}(u,v)}$, $a=1$      Fig. 5: $\dfrac{c^{NB}_{\mathbf{a},\text{sing}}(u,v)}{c^{\Gamma}_{\mathbf{a},\text{sing}}(u,v)}$, $a=3$

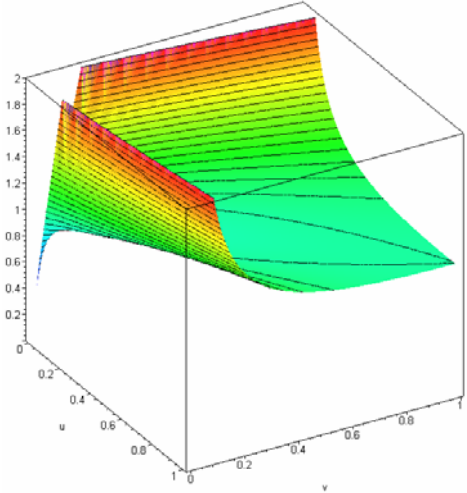 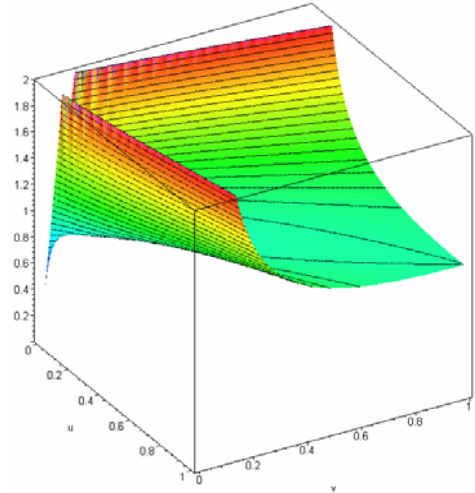

Fig. 6: $\dfrac{c^{NB}_{\mathbf{a},\text{sing}}(u,v)}{c^{\Gamma}_{\mathbf{a},\text{sing}}(u,v)}$, $a=7$      Fig. 7: $\dfrac{c^{NB}_{a,\text{sing}}(u,v)}{c^{\Gamma}_{a,\text{sing}}(u,v)}$, $a=10$

The main difference between both copulas obviously lies in the lower south-west corner of the unit square.

In what follows we consider the usual upper and lower tails dependence coefficients $\lambda_U$ and $\lambda_L$ as described e.g. in McNeil et al. [4], chapter 7.2.4. For the proof of Theorem 2, the following result will be needed.

**Lemma.** For $m \in \mathbb{N}$, there holds $\sum_{k=0}^{m} \binom{m+k-1}{k} \dfrac{1}{2^k} = \dfrac{1}{2^{m+1}} \left[ \binom{2m}{m} + 4^m \right]$.

*Proof.* In Stanley [9], 5.53, p.144 the following relation can be found:

(24) $$\sum_{j=0}^{m-1}\binom{m+j-1}{j}\frac{1}{2^j}=2^{m-1},$$

which directly implies

(25) $$\sum_{j=0}^{m}\binom{m+j-1}{j}\frac{1}{2^j}=2^{m-1}+\frac{1}{2^m}\binom{2m-1}{m}=2^{m-1}+\frac{1}{2^{m+1}}\binom{2m}{m}=\frac{1}{2^{m+1}}\left[\binom{2m}{m}+4^m\right]. \blacksquare$$

**Theorem 2.** The upper tail dependence coefficient $\lambda_U(\mathbf{a})$ of the diagonal dominant Gamma copula is identical to that of a diagonal dominant negative binomial copula $c^{NB}$ for $a \in \mathbb{N}$, given by

(26) $$\lambda_U(\mathbf{a}):=\lim_{t\to 1}\frac{\int_t^1\int_t^1 c_{\mathbf{a},\text{sing}}^{\Gamma}(u,v)\,du\,dv}{1-t}=a\int_0^{\infty}\left(\frac{1-e^{-z}\sum_{k=0}^{a}\frac{z^k}{k!}}{z}\right)^2 dz$$

$$=1-\frac{1}{4^a}\binom{2a}{a}\sim 1-\frac{1}{\sqrt{\pi a}}\quad(a\to\infty).$$

*Proof.* Let $F_a(x):=1-e^{-x}\sum_{k=0}^{a-1}\frac{x^k}{k!}$, $x>0$ denote the distribution function of the standard Gamma distribution with shape parameter $a \in \mathbb{N}$. It holds

(27) $$J_a(t):=\int_t^1\int_t^1 c_{\mathbf{a},\text{sing}}^{\Gamma}(u,v)\,du\,dv=a\int_0^{\infty}\frac{s^{a-1}(1+s)^{a+1}}{\Gamma^2(a+1)}\int_t^1\int_t^1(\ln u\cdot\ln v)^a\cdot u^s\cdot v^s\,du\,dv\,ds$$

$$=a\int_0^{\infty}s^{a-1}(1+s)^{a+1}\left[\int_t^1\frac{(-\ln w)^a}{\Gamma(a+1)}\cdot w^s\,dw\right]^2 ds.$$

Now by the substitution $-\ln w = x$ or $w = e^{-x}$ with $dw = -e^{-x}dx$ we get the expression

$$\int_t^1\frac{(-\ln w)^a}{\Gamma(a+1)}\cdot w^s\,dw=\int_0^{-\ln t}\frac{x^a}{\Gamma(a+1)}e^{-(1+s)x}\,dx=\frac{1}{(1+s)^{a+1}}\int_0^{-\ln t}(1+s)^{a+1}\frac{x^a}{\Gamma(a+1)}e^{-(1+s)x}\,dx$$

$$=\frac{1}{(1+s)^{a+1}}F_{a+1}\big((1+s)\cdot(-\ln t)\big)$$

and hence

$$J_a(t)=a\int_0^{\infty}\frac{s^{a-1}}{(1+s)^{a+1}}F_{a+1}^2\big((1+s)\cdot(-\ln t)\big)\,ds.$$

Substituting again $(1+s)\cdot(-\ln t)=z$, we obtain

(28) $$J_a(t)=(-\ln t)\cdot a\cdot\int_{-\ln t}^{\infty}\frac{(z+\ln t)^{a-1}}{z^{a+1}}F_{a+1}^2(z)\,dz.$$

This implies

$$(29) \quad \lambda_U(a) = \lim_{t \uparrow 1} \frac{J_a(t)}{1-t} = a \lim_{t \uparrow 1} \frac{-\ln t}{1-t} \cdot \lim_{t \uparrow 1} \int_{-\ln t}^{\infty} \frac{(z + \ln t)^{a-1}}{z^{a+1}} F_{a+1}^2(z)\, dz = a \int_0^{\infty} \frac{F_{a+1}^2(z)}{z^2}\, dz$$

which proves the first equality above. For the remainder, we proceed by induction on $a$. The case $a = 1$ is evident because of

$$(30) \quad \int_0^{\infty} \frac{F_2^2(z)}{z^2}\, dz = -\left[\frac{(1-e^{-x})^2}{x} + \frac{e^{-2x}}{2}\right]_0^{\infty} = \frac{1}{2} = 1 - \frac{1}{4}\binom{2}{1}.$$

For $a \geq 2$, assume that the equality $(a-1)\int_0^{\infty} \frac{F_a^2(z)}{z^2}\, dz = 1 - \frac{1}{4^{a-1}}\binom{2a-2}{a-1}$ holds. Since

$$F_{a+1}^2(z) = \left\{\left(1 - e^{-z}\sum_{k=0}^{a-1}\frac{z^k}{k!}\right) - \frac{z^a}{a!}e^{-z}\right\}^2 = F_a^2(z) + \left(\frac{z^a}{a!}e^{-z}\right)^2 - 2F_a(z)\frac{z^a}{a!}e^{-z}$$

$$= F_a^2(z) + \frac{z^{2a}}{(a!)^2}e^{-2z} - 2\frac{z^a}{a!}e^{-z} + 2\frac{z^a}{a!}e^{-2z}\sum_{k=0}^{a-1}\frac{z^k}{k!},$$

we obtain

$$a\int_0^{\infty} \frac{F_{a+1}^2(z)}{z^2}\, dz = a\left(\int_0^{\infty} \frac{F_a^2(z)}{z^2}\, dz + \frac{1}{(a!)^2}\int_0^{\infty} z^{2a-2}e^{-2z}\, dz - \frac{2}{a!}\int_0^{\infty} z^{a-2}e^{-z}\, dz + \frac{2}{a!}\int_0^{\infty} z^{a-2}e^{-2z}\sum_{k=0}^{a-1}\frac{z^k}{k!}\, dz\right).$$

An easy computation shows that

$$\int_0^{\infty} z^{2a-2}e^{-2z}\, dz = \frac{(2a-2)!}{2^{2a-1}}\int_0^{\infty} \frac{2^{2a-1}}{(2a-2)!}z^{2a-2}e^{-2z}\, dz = \frac{(2a-2)!}{2^{2a-1}},$$

$$\int_0^{\infty} z^{a-2}e^{-z}\, dz = (a-2)!\int_0^{\infty} \frac{1}{(a-2)!}z^{a-2}e^{-z}\, dz = (a-2)! \quad \text{and, likewise,}$$

$$\int_0^{\infty} z^{a-2}e^{-2z}\sum_{k=0}^{a-1}\frac{z^k}{k!}\, dz = \sum_{k=0}^{a-1}\frac{1}{k!}\int_0^{\infty} z^{a+k-2}e^{-2z}\, dz = \sum_{k=0}^{a-1}\frac{(a+k-2)!}{k!\, 2^{a+k-1}}.$$

Substituting this in the expression above, we get, with the Lemma above,

$$a\int_0^{\infty} \frac{F_{a+1}^2(z)}{z^2}\, dz = \frac{a}{a-1}\left\{1 - \frac{1}{4^{a-1}}\binom{2a-2}{a-1}\right\} + \frac{(2a-2)!}{a!(a-1)!\, 2^{2a-1}} - \frac{2(a-2)!}{(a-1)!} + \frac{1}{(a-1)!}\sum_{k=0}^{a-1}\frac{(a+k-2)!}{k!\, 2^{a+k-2}}$$

$$= \frac{a}{a-1}\left\{1 - \frac{1}{4^{a-1}}\binom{2a-2}{a-1}\right\} + \frac{1}{2a \cdot 4^{a-1}}\binom{2a-2}{a-1} - \frac{2}{a-1} + \frac{1}{(a-1)2^{a-2}}\frac{1}{2^a}\left[\binom{2a-2}{a-1} + 4^{a-1}\right]$$

$$= 1 - \frac{2a-1}{2a}\frac{1}{4^{a-1}}\binom{2a-2}{a-1} = 1 - \frac{1}{4^a}\binom{2a}{a},$$

i.e. the statement is also true for $a+1$. This proves the assertion. ∎

Fig. 8 to Fig. 11 show some symmetric Gamma copula densities $c_{\mathbf{a},\text{sing}}^{\Gamma}(u,v)$ for different values of $a$.

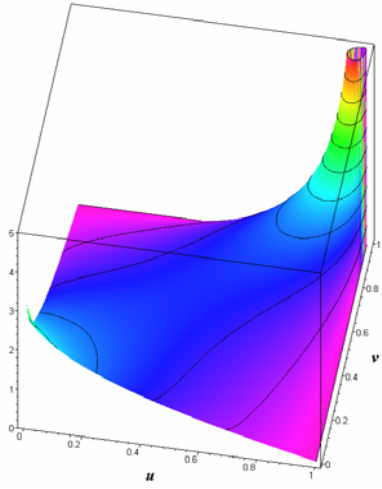

Fig. 8: $c_{\mathbf{a},\text{sing}}^{\Gamma}$, $a=1$

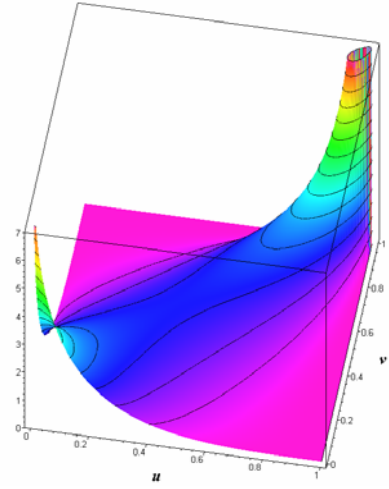

Fig. 9: $c_{\mathbf{a},\text{sing}}^{\Gamma}$, $a=3$

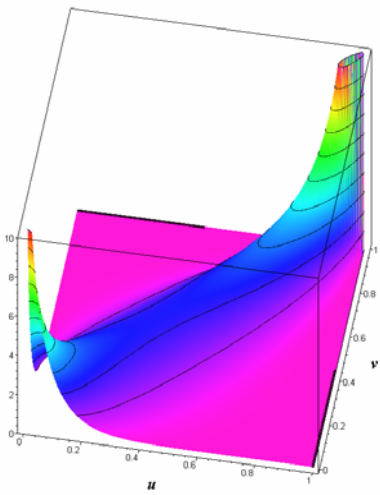

Fig. 10: $c_{\mathbf{a},\text{sing}}^{\Gamma}$, $a=8$

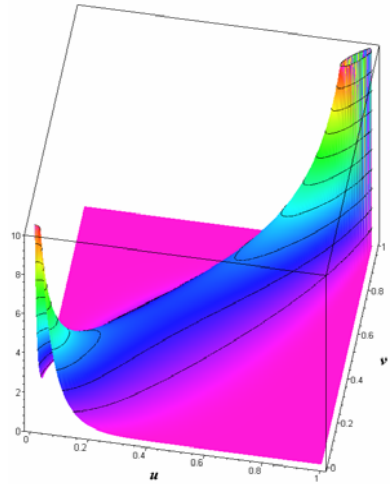

Fig. 11: $c_{\mathbf{a},\text{sing}}^{\Gamma}$, $a=9$

In contrast, the Power copula does not show a tail dependence, no matter what the parameters are.

**Theorem 3.** The upper and lower tail dependence coefficients $\lambda_U(\beta)$ and $\lambda_L(\beta)$ of the diagonal dominant Power copula are zero for all $\beta > 2$.

*Proof.* Due to symmetry, it suffices to prove the theorem for $\lambda_U(\beta)$ alone. Like in the proof of theorem 2 (see relations (27) and (29)), we have

$$(31) \quad \lambda_U(\beta) := \lim_{t \to 1} \frac{1}{1-t} \int_0^1 \frac{1}{\alpha(s)} \left[ \int_t^1 \varphi(s,u)\,du \right]^2 ds =: \lim_{t \to 1} \int_0^1 K(s,t)\,ds$$

with $K(s,t) := \dfrac{1}{\alpha(s)(1-t)} \left[ \int_t^1 \varphi(s,u)\,du \right]^2$ where

$$\text{(32)} \quad \int_t^1 \varphi(s,u)\,du = \frac{\beta}{\beta-2}\begin{cases} \dfrac{s^{\beta-1}}{t^{\beta-2}}\left(1-t^{\beta-2}\right), & s \le t, \\[6pt] 1-s^{\beta-1}-\dfrac{(1-s)^{\beta-1}}{(1-t)^{\beta-2}}, & s > t, \end{cases} \quad 0 < s,t < 1.$$

It is easy to see that $K(s,t)$ as a function of $s$ attains its maximum at $s = t$ with

$$\text{(33)} \quad K(t,t) = \frac{\beta}{\beta-2} \cdot \frac{t^2\left(1-t^{\beta-2}\right)^2}{t^{2(\beta-2)}\left(1-t^{\beta-1}-(1-t)^{\beta-1}\right)(1-t)} \le \frac{\beta}{\beta-2} \cdot \lim_{t\to 1}\frac{t^2\left(1-t^{\beta-2}\right)^2}{\left(1-t^{\beta-1}-(1-t)^{\beta-1}\right)(1-t)}$$

$$= \frac{\beta}{\beta-2}\cdot\frac{(\beta-2)^2}{\beta-1} = \frac{\beta(\beta-2)}{\beta-1}$$

for $\beta \ge 3$ from where we get

$$\text{(34)} \quad \lim_{t\to 1}\int_t^1 K(s,t)\,ds \le \lim_{t\to 1}\int_t^1 \frac{\beta(\beta-2)}{\beta-1}\,ds = 0.$$

A simple analysis shows that for $2 < \beta < 3$, $K(t,t)$ also is bounded, by the constant ½, so that the limit relation (31) is valid for all $\beta > 2$. It remains to show that $\lim_{t\to 1}\int_0^t K(s,t)\,ds = 0$. For this purpose, we use the elementary estimate

$$\text{(35)} \quad \frac{1}{1-s^{\beta-1}-(1-s)^{\beta-1}} \le \frac{1}{2-\dfrac{1}{2^{\beta-3}}}\cdot \max\left(\frac{1}{s},\frac{1}{1-s}\right) \text{ for } 0 < s < 1.$$

We thus obtain, for $t > $ ½,

$$\text{(36)} \quad \int_0^t K(s,t)\,ds = \int_0^{1/2} K(s,t)\,ds + \int_{1/2}^t K(s,t)\,ds \le \frac{\beta}{\beta-2}\cdot\frac{1}{2-\dfrac{1}{2^{\beta-3}}}\cdot\frac{\left(1-t^{\beta-2}\right)^2}{t^{2(\beta-2)}(1-t)}\left[\int_0^{1/2} s^{2\beta-3}\,ds + \int_{1/2}^t \frac{s^{2\beta-2}}{1-s}\,ds\right]$$

$$\le \frac{\beta}{\beta-2}\cdot\frac{1}{2-\dfrac{1}{2^{\beta-3}}}\cdot\frac{\left(1-t^{\beta-2}\right)^2}{t^{2(\beta-2)}(1-t)}\left[\frac{1}{2^{2\beta-1}(\beta-1)} - t^{2\beta-2}\ln\left(\frac{1-t}{2}\right)\right].$$

The final result now follows by the observation

$$\text{(37)} \quad \lim_{t\to 1}\frac{\left(1-t^{\beta-2}\right)^2}{(1-t)}\ln(1-t) = g'(1) = 0$$

for the function $g(t) = \left(1-t^{\beta-2}\right)^2\ln(1-t)$, $0 < t < 1$. ∎

Fig. 12 and Fig. 13 show the Power copula densities $c_\beta^P(u,v)$ in the diagonal dominant case for different values of $\beta$.

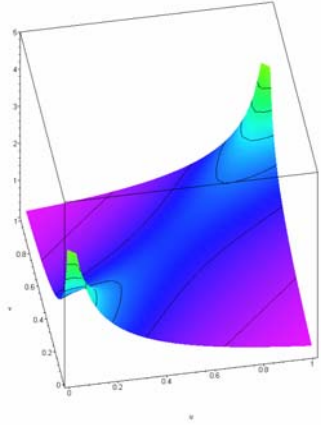 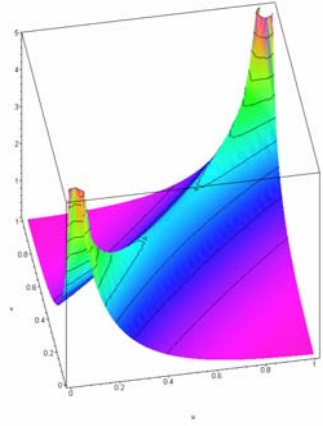

Fig. 12: $c_3^P(u,v)$      Fig. 13: $c_5^P(u,v)$

## 5 Case Study A

Firstly, we extend the example data set given in Cottin and Pfeifer [1] because it was also used as a data basis in several former papers on partition-of-unity-copulas (Pfeifer et al. [6], [7]). $\mathbf{r}_1$ and $\mathbf{r}_2$ denote the correspondig rank vectors which are the basis for the empirical copula and different copula drivers, as described in section 2 above.

| no. | risk $X_1$ | risk $X_2$ | $\mathbf{r}_1$ | $\mathbf{r}_2$ |
|---|---|---|---|---|
| 1 | 0.468 | 0.966 | 4 | 9 |
| 2 | 9.951 | 2.679 | 20 | 20 |
| 3 | 0.866 | 0.897 | 8 | 4 |
| 4 | 6.731 | 2.249 | 19 | 19 |
| 5 | 1.421 | 0.956 | 13 | 8 |
| 6 | 2.040 | 1.141 | 17 | 15 |
| 7 | 2.967 | 1.707 | 18 | 18 |
| 8 | 1.200 | 1.008 | 11 | 10 |
| 9 | 0.426 | 1.065 | 3 | 12 |
| 10 | 1.946 | 1.162 | 15 | 16 |
| 11 | 0.676 | 0.918 | 5 | 6 |
| 12 | 1.184 | 1.336 | 10 | 17 |
| 13 | 0.960 | 0.933 | 9 | 7 |
| 14 | 1.972 | 1.077 | 16 | 13 |
| 15 | 1.549 | 1.041 | 14 | 11 |
| 16 | 0.819 | 0.899 | 6 | 5 |
| 17 | 0.063 | 0.710 | 1 | 1 |
| 18 | 1.280 | 1.118 | 12 | 14 |
| 19 | 0.824 | 0.894 | 7 | 3 |
| 20 | 0.227 | 0.837 | 2 | 2 |

Tab. 1: the data from Cottin and Pfeifer

Fig. 14 to Fig. 19 show some simulated examples for a selection of copula drivers based on the multivariate normal distribution, with different choices of the correlation parameter $\rho \in [-1,1]$.

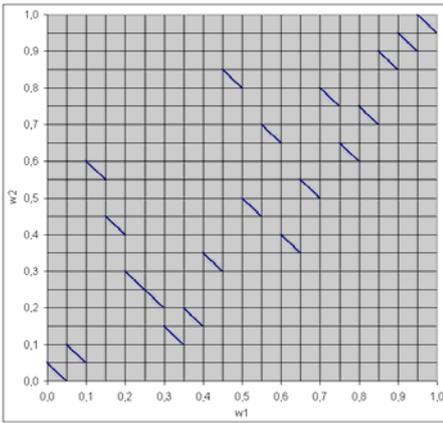
LF copula driver, $\rho = -1$

Fig. 14

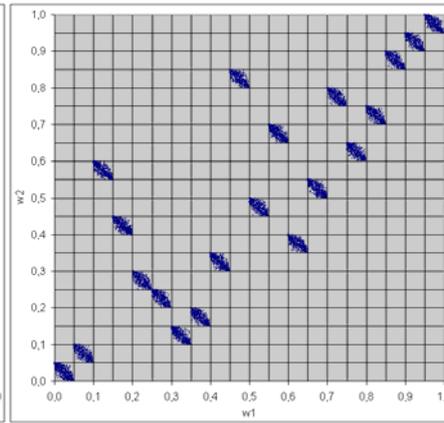
normal copula driver, $\rho = -0.8$

Fig. 15

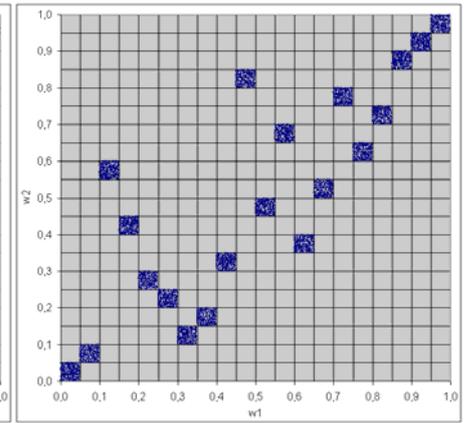
rook copula driver, $\rho = 0$

Fig. 16

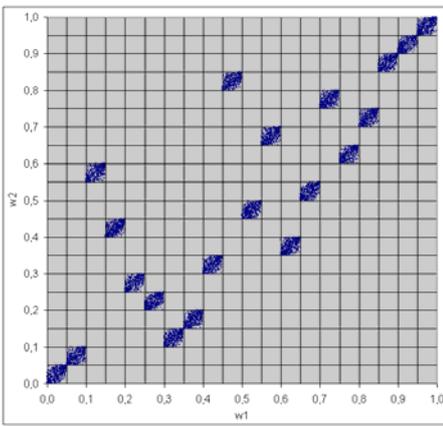
normal copula driver, $\rho = 0.6$

Fig. 17

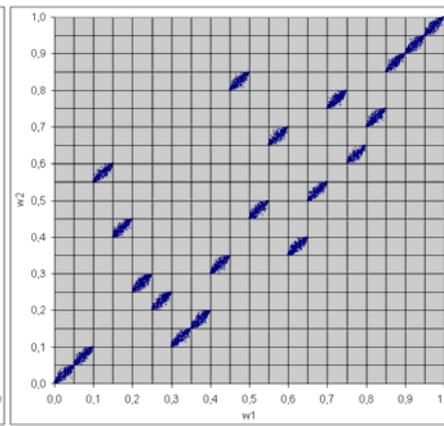
normal copula driver, $\rho = 0.9$

Fig. 18

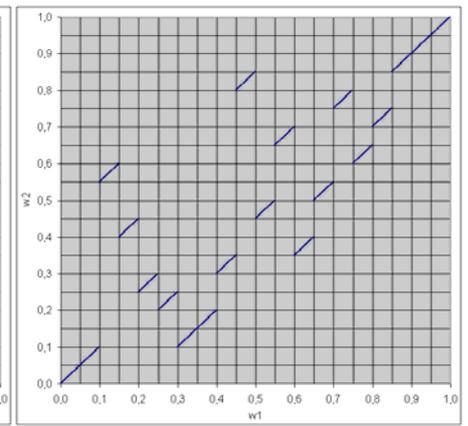
UF copula driver, $\rho = 1$

Fig. 19

Fig. 20 to Fig. 25 show some Monte Carlo realizations for the Gamma copula on the basis of the data set together with the empirical copula (relative rank vectors: circular points).

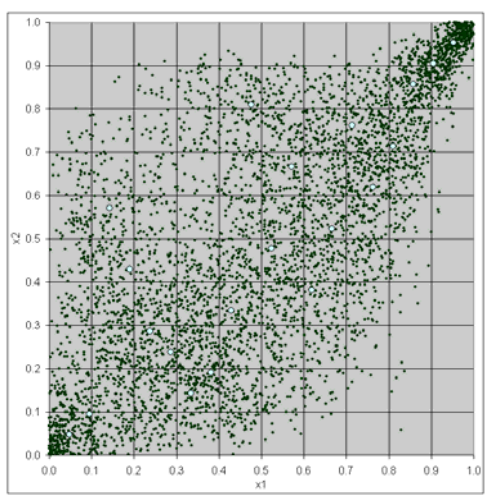
with rook copula driver, $a_1 = a_2 = 7$

Fig. 20

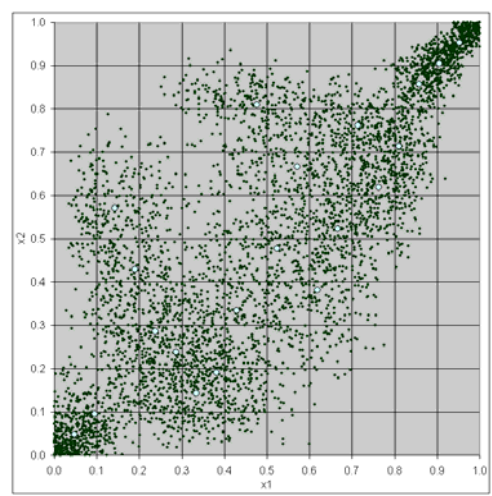
with rook copula driver, $a_1 = a_2 = 15$

Fig. 21

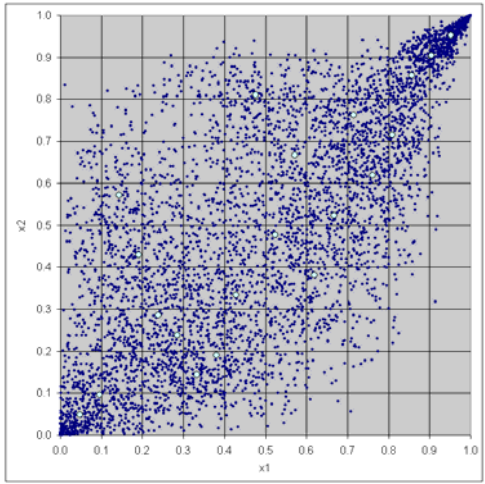 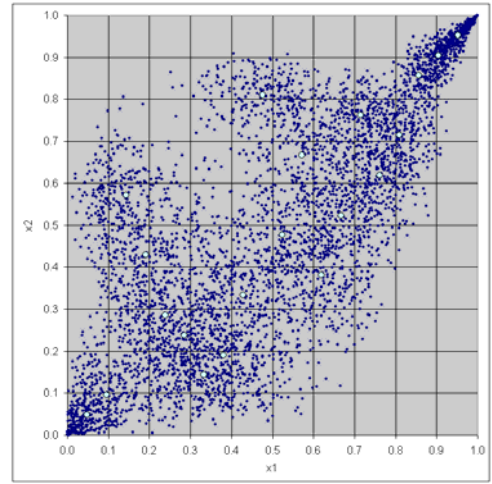

| with UF copula driver, $a_1 = a_2 = 7$ | with UF copula driver, $a_1 = a_2 = 15$ |
|---|---|
| Fig. 22 | Fig. 23 |

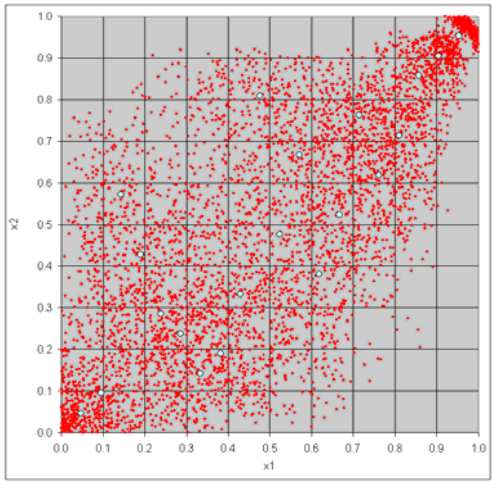 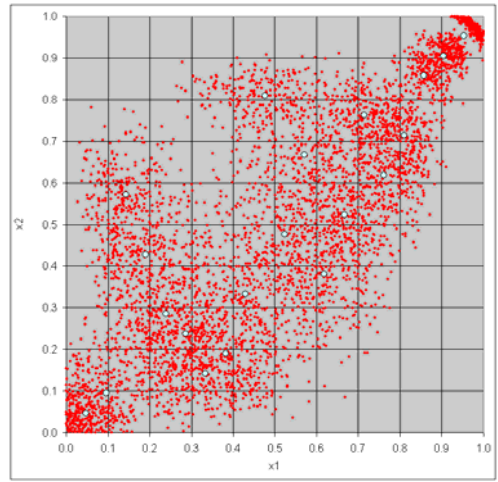

| with LF copula driver, $a_1 = a_2 = 7$ | with LF copula driver, $a_1 = a_2 = 15$ |
|---|---|
| Fig. 24 | Fig. 25 |

For the Power copula, the following graphs show the results of a corresponding Monte Carlo study.

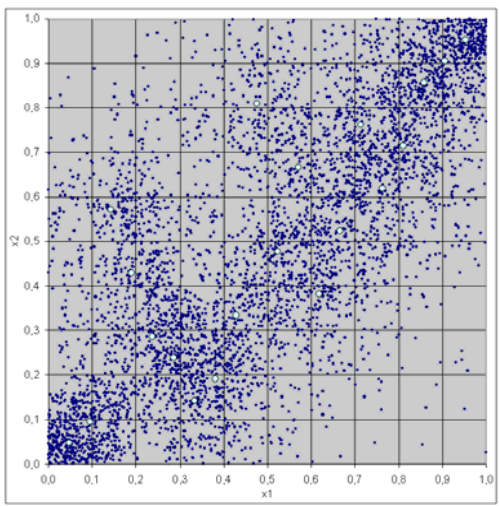 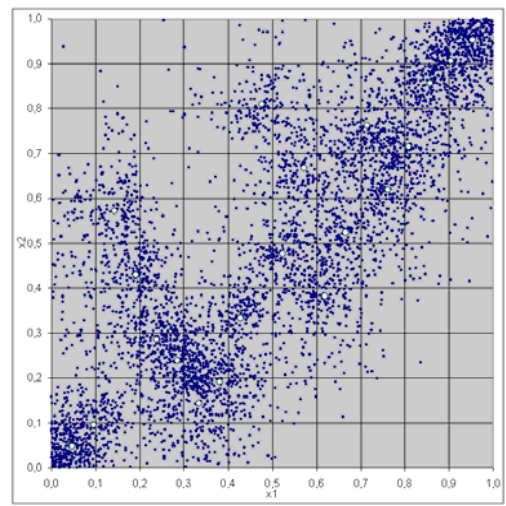

| with rook copula driver, $\beta_1 = \beta_2 = 8$ | with rook copula driver, $\beta_1 = \beta_2 = 12$ |
|---|---|
| Fig. 26 | Fig. 27 |

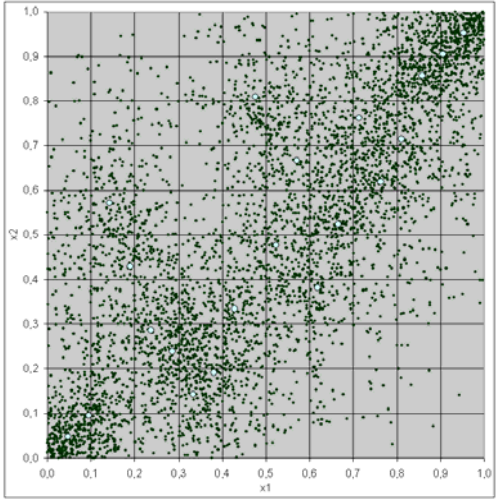

with UF copula driver, $\beta_1 = \beta_2 = 8$

Fig. 28

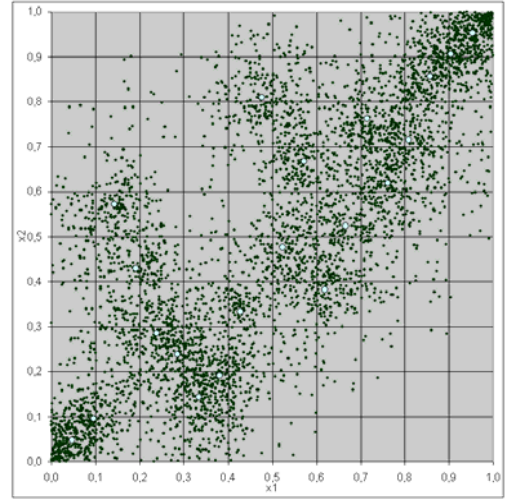

with UF copula driver, $\beta_1 = \beta_2 = 12$

Fig. 29

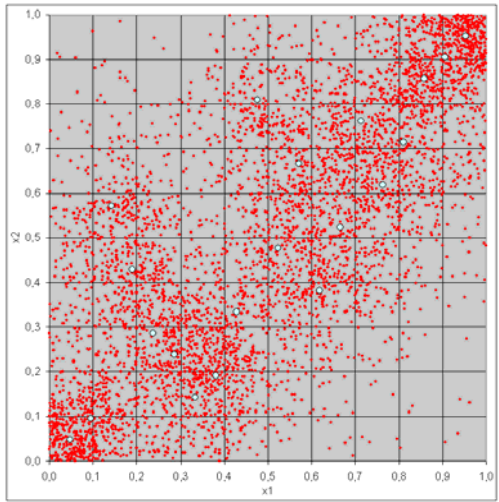

with LF copula driver, $\beta_1 = \beta_2 = 8$

Fig. 30

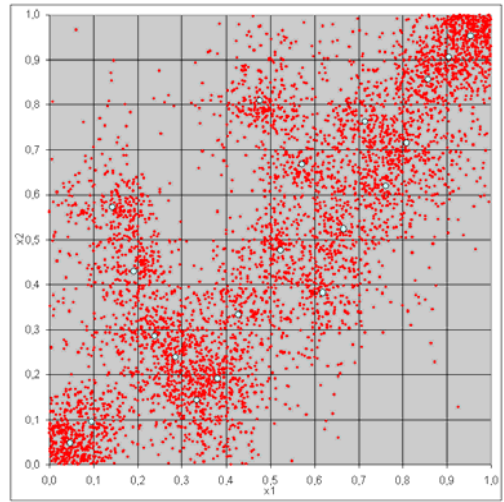

with LF copula driver, $\beta_1 = \beta_2 = 12$

Fig. 31

A comparison of Fig. 22 and Fig. 23 with Fig. 28 and Fig. 29 indicates also visually that the Power copula possesses no tail dependence.

**6 Case Study B**

In order to show the powerfulness of continuous PUC approaches in higher dimensions we conclude the applied section with a discussion of the 19-dimensional data set presented in Neumann et al. [5], listed in Tab. 2 and Tab. 3, containing insurance losses from a non-life portfolio of natural perils in 19 areas in central Europe over a time period of 20 years. The monetary unit is 1 million €

For simplicity, we will consider only the Gamma copula in this section.

| Year | Area 1 | Area 2 | Area 3 | Area 4 | Area 5 | Area 6 | Area 7 | Area 8 | Area 9 | Area 10 |
|---|---|---|---|---|---|---|---|---|---|---|
| 1 | 23.664 | 154.664 | 40.569 | 14.504 | 10.468 | 7.464 | 22.202 | 17.682 | 12.395 | 18.551 |
| 2 | 1.080 | 59.545 | 3.297 | 1.344 | 1.859 | 0.477 | 6.107 | 7.196 | 1.436 | 3.720 |
| 3 | 21.731 | 31.049 | 55.973 | 5.816 | 14.869 | 20.771 | 3.580 | 14.509 | 17.175 | 87.307 |
| 4 | 28.99 | 31.052 | 30.328 | 4.709 | 0.717 | 3.530 | 6.032 | 6.512 | 0.682 | 3.115 |
| 5 | 53.616 | 62.027 | 57.639 | 1.804 | 2.073 | 4.361 | 46.018 | 22.612 | 1.581 | 11.179 |
| 6 | 29.95 | 41.722 | 12.964 | 1.127 | 1.063 | 4.873 | 6.571 | 11.966 | 15.676 | 24.263 |
| 7 | 3.474 | 14.429 | 10.869 | 0.945 | 2.198 | 1.484 | 4.547 | 2.556 | 0.456 | 1.137 |
| 8 | 10.02 | 31.283 | 21.116 | 1.663 | 2.153 | 0.932 | 25.163 | 3.222 | 1.581 | 5.477 |
| 9 | 5.816 | 14.804 | 128.072 | 0.523 | 0.324 | 0.477 | 3.049 | 7.791 | 4.079 | 7.002 |
| 10 | 170.725 | 576.767 | 108.361 | 41.599 | 20.253 | 35.412 | 126.698 | 71.079 | 21.762 | 64.582 |
| 11 | 21.423 | 50.595 | 4.360 | 0.327 | 1.566 | 64.621 | 5.650 | 1.258 | 0.626 | 3.556 |
| 12 | 6.38 | 28.316 | 3.740 | 0.442 | 0.736 | 0.470 | 3.406 | 7.859 | 0.894 | 3.591 |
| 13 | 124.665 | 33.359 | 14.712 | 0.321 | 0.975 | 2.005 | 3.981 | 4.769 | 2.006 | 1.973 |
| 14 | 20.165 | 49.948 | 17.658 | 0.595 | 0.548 | 29.35 | 6.782 | 4.873 | 2.921 | 6.394 |
| 15 | 78.106 | 41.681 | 13.753 | 0.585 | 0.259 | 0.765 | 7.013 | 9.426 | 2.18 | 3.769 |
| 16 | 11.067 | 444.712 | 365.351 | 99.366 | 8.856 | 28.654 | 10.589 | 13.621 | 9.589 | 19.485 |
| 17 | 6.704 | 81.895 | 14.266 | 0.972 | 0.519 | 0.644 | 8.057 | 18.071 | 5.515 | 13.163 |
| 18 | 15.55 | 277.643 | 26.564 | 0.788 | 0.225 | 1.230 | 26.800 | 64.538 | 2.637 | 80.711 |
| 19 | 10.099 | 18.815 | 9.352 | 2.051 | 1.089 | 6.102 | 2.678 | 4.064 | 2.373 | 2.057 |
| 20 | 8.492 | 138.708 | 46.708 | 3.68 | 1.132 | 1.698 | 165.6 | 7.926 | 2.972 | 5.237 |

Tab. 2: loss data, part one

| Year | Area 11 | Area 12 | Area 13 | Area 14 | Area 15 | Area 16 | Area 17 | Area 18 | Area 19 |
|---|---|---|---|---|---|---|---|---|---|
| 1 | 1.842 | 4.100 | 46.135 | 14.698 | 44.441 | 7.981 | 35.833 | 10.689 | 7.299 |
| 2 | 0.429 | 1.026 | 7.469 | 7.058 | 4.512 | 0.762 | 14.474 | 9.337 | 0.740 |
| 3 | 0.209 | 2.344 | 22.651 | 4.117 | 26.586 | 3.920 | 13.804 | 2.683 | 3.026 |
| 4 | 0.521 | 0.696 | 31.126 | 1.878 | 29.423 | 6.394 | 18.064 | 1.201 | 0.894 |
| 5 | 2.715 | 1.327 | 40.156 | 4.655 | 104.691 | 28.579 | 17.832 | 1.618 | 3.402 |
| 6 | 4.832 | 0.701 | 16.712 | 11.852 | 29.234 | 7.098 | 17.866 | 5.206 | 5.664 |
| 7 | 0.268 | 0.580 | 11.851 | 2.057 | 11.605 | 0.282 | 16.925 | 2.082 | 1.008 |
| 8 | 0.741 | 0.369 | 3.814 | 1.869 | 8.126 | 1.032 | 14.985 | 1.390 | 1.703 |
| 9 | 0.524 | 6.554 | 5.459 | 3.007 | 8.528 | 1.920 | 5.638 | 2.149 | 2.908 |
| 10 | 9.882 | 6.401 | 106.197 | 44.912 | 191.809 | 90.559 | 154.492 | 36.626 | 36.276 |
| 11 | 1.052 | 8.277 | 22.564 | 8.961 | 19.817 | 16.437 | 25.990 | 2.364 | 6.434 |
| 12 | 0.136 | 0.364 | 28.000 | 7.574 | 3.213 | 1.749 | 12.735 | 1.744 | 0.558 |
| 13 | 1.990 | 15.176 | 57.235 | 23.686 | 110.035 | 17.373 | 7.276 | 2.494 | 0.525 |
| 14 | 0.630 | 0.762 | 25.897 | 3.439 | 8.161 | 3.327 | 24.733 | 2.807 | 1.618 |
| 15 | 0.770 | 15.024 | 36.068 | 1.613 | 6.127 | 8.103 | 12.596 | 4.894 | 0.822 |
| 16 | 0.287 | 0.464 | 24.211 | 38.616 | 51.889 | 1.316 | 173.080 | 3.557 | 11.627 |
| 17 | 0.590 | 2.745 | 16.124 | 2.398 | 20.997 | 2.515 | 5.161 | 2.840 | 3.002 |
| 18 | 0.245 | 0.217 | 12.416 | 4.972 | 59.417 | 3.762 | 24.603 | 7.404 | 19.107 |
| 19 | 0.415 | 0.351 | 10.707 | 2.468 | 10.673 | 1.743 | 27.266 | 1.368 | 0.644 |
| 20 | 0.566 | 0.708 | 22.646 | 6.652 | 14.437 | 63.692 | 113.231 | 7.218 | 2.548 |

Tab.3: loss data, part two

As is to be expected, insurance losses in locally adjacent areas show a high degree of stochastic dependence, which can also be seen from the following empirical correlation tables (Tab. 5). For a better readability, only two decimal places are displayed. Correlation coefficients above 90% are highlighted.

|     | A1 | A2 | A3 | A4 | A5 | A6 | A7 | A8 | A9 | A10 | A11 | A12 | A13 | A14 | A15 | A16 | A17 | A18 | A19 |
|-----|----|----|----|----|----|----|----|----|----|-----|-----|-----|-----|-----|-----|-----|-----|-----|-----|
| A1  | 1 | 0.46 | 0.03 | 0.16 | 0.47 | 0.20 | 0.35 | 0.49 | 0.41 | 0.24 | 0.78 | 0.64 | 0.91 | 0.63 | 0.85 | 0.66 | 0.30 | 0.67 | 0.56 |
| A2  | 0.46 | 1 | 0.64 | 0.78 | 0.67 | 0.36 | 0.51 | 0.76 | 0.57 | 0.51 | 0.58 | -0.04 | 0.59 | 0.84 | 0.68 | 0.58 | 0.87 | 0.77 | 0.90 |
| A3  | 0.03 | 0.64 | 1 | 0.93 | 0.41 | 0.26 | 0.11 | 0.16 | 0.33 | 0.16 | 0.08 | -0.09 | 0.13 | 0.64 | 0.25 | 0.10 | 0.74 | 0.14 | 0.35 |
| A4  | 0.16 | 0.78 | 0.93 | 1 | 0.54 | 0.36 | 0.16 | 0.25 | 0.43 | 0.19 | 0.22 | -0.10 | 0.30 | 0.79 | 0.36 | 0.19 | 0.84 | 0.32 | 0.49 |
| A5  | 0.47 | 0.67 | 0.41 | 0.54 | 1 | 0.41 | 0.35 | 0.51 | 0.84 | 0.63 | 0.59 | 0.02 | 0.64 | 0.67 | 0.59 | 0.50 | 0.58 | 0.71 | 0.67 |
| A6  | 0.20 | 0.36 | 0.26 | 0.36 | 0.41 | 1 | 0.07 | 0.11 | 0.28 | 0.19 | 0.28 | 0.14 | 0.31 | 0.42 | 0.24 | 0.27 | 0.39 | 0.27 | 0.40 |
| A7  | 0.35 | 0.51 | 0.11 | 0.16 | 0.35 | 0.07 | 1 | 0.44 | 0.27 | 0.19 | 0.48 | -0.07 | 0.46 | 0.35 | 0.45 | 0.91 | 0.64 | 0.61 | 0.49 |
| A8  | 0.49 | 0.76 | 0.16 | 0.25 | 0.51 | 0.11 | 0.44 | 1 | 0.50 | 0.75 | 0.61 | -0.03 | 0.54 | 0.47 | 0.71 | 0.53 | 0.40 | 0.75 | 0.90 |
| A9  | 0.41 | 0.57 | 0.33 | 0.43 | 0.84 | 0.28 | 0.27 | 0.50 | 1 | 0.66 | 0.68 | -0.01 | 0.52 | 0.60 | 0.50 | 0.41 | 0.46 | 0.65 | 0.63 |
| A10 | 0.24 | 0.51 | 0.16 | 0.19 | 0.63 | 0.19 | 0.19 | 0.75 | 0.66 | 1 | 0.33 | -0.12 | 0.27 | 0.28 | 0.43 | 0.24 | 0.23 | 0.45 | 0.65 |
| A11 | 0.78 | 0.58 | 0.08 | 0.22 | 0.59 | 0.28 | 0.48 | 0.61 | 0.68 | 0.33 | 1 | 0.19 | 0.79 | 0.65 | 0.80 | 0.73 | 0.43 | 0.84 | 0.74 |
| A12 | 0.64 | -0.04 | -0.09 | -0.10 | 0.02 | 0.14 | -0.07 | -0.03 | -0.01 | -0.12 | 0.19 | 1 | 0.44 | 0.21 | 0.28 | 0.17 | -0.12 | 0.13 | 0.03 |
| A13 | 0.91 | 0.59 | 0.13 | 0.30 | 0.64 | 0.31 | 0.46 | 0.54 | 0.52 | 0.27 | 0.79 | 0.44 | 1 | 0.71 | 0.86 | 0.74 | 0.47 | 0.76 | 0.65 |
| A14 | 0.63 | 0.84 | 0.64 | 0.79 | 0.67 | 0.42 | 0.35 | 0.47 | 0.60 | 0.28 | 0.65 | 0.21 | 0.71 | 1 | 0.74 | 0.54 | 0.79 | 0.68 | 0.72 |
| A15 | 0.85 | 0.68 | 0.25 | 0.36 | 0.59 | 0.24 | 0.45 | 0.71 | 0.50 | 0.43 | 0.80 | 0.28 | 0.86 | 0.74 | 1 | 0.69 | 0.47 | 0.71 | 0.75 |
| A16 | 0.66 | 0.58 | 0.10 | 0.19 | 0.50 | 0.27 | 0.91 | 0.53 | 0.41 | 0.24 | 0.73 | 0.17 | 0.74 | 0.54 | 0.69 | 1 | 0.63 | 0.77 | 0.64 |
| A17 | 0.30 | 0.87 | 0.74 | 0.84 | 0.58 | 0.39 | 0.64 | 0.40 | 0.46 | 0.23 | 0.43 | -0.12 | 0.47 | 0.79 | 0.47 | 0.63 | 1 | 0.59 | 0.64 |
| A18 | 0.67 | 0.77 | 0.14 | 0.32 | 0.71 | 0.27 | 0.61 | 0.75 | 0.65 | 0.45 | 0.84 | 0.13 | 0.76 | 0.68 | 0.71 | 0.77 | 0.59 | 1 | 0.86 |
| A19 | 0.56 | 0.90 | 0.35 | 0.49 | 0.67 | 0.40 | 0.49 | 0.90 | 0.63 | 0.65 | 0.74 | 0.03 | 0.65 | 0.72 | 0.75 | 0.64 | 0.64 | 0.86 | 1 |

Tab 4: empirical correlations between original losses in adjacent areas

For a comparison of copula models, we have used classical approaches with a Gaussian and a *t*-copula (with two degrees of freedom for modelling a high degree of tail dependence), as well as with a Gamma copula for different choices of the copula drivers (rook and UF) with $a_k = 10$ for $k = 1, \cdots, 19$. The graphs displayed in Fig. 32 to Fig. 47 show a selection of the 171 possible pairwise two-dimensional projections of corresponding Monte Carlo simulations $U_k$ where the highest pairwise correlations have been observed, together with the empirical copulas (relative rank vectors: circular points). The parameter matrices for the Gaussian and *t*-copulas were calculated from the empirical correlations of log data.

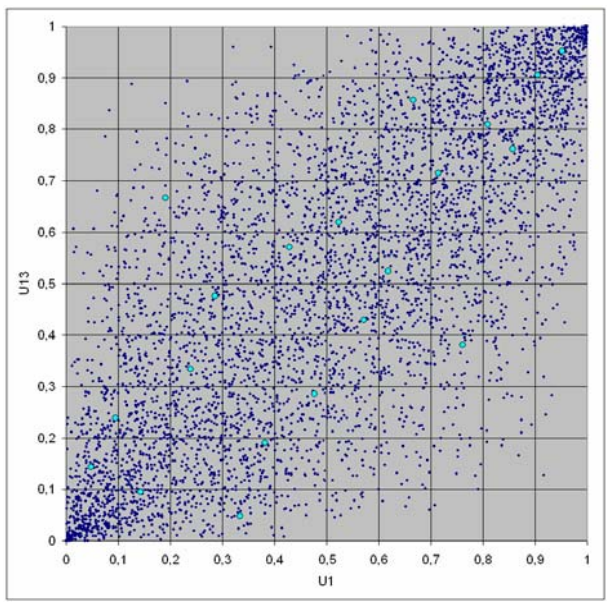
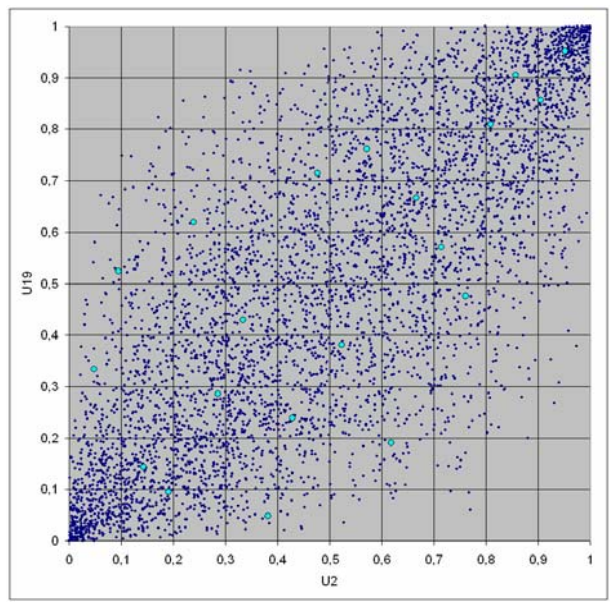

Fig. 32      Fig. 33

Gaussian copula

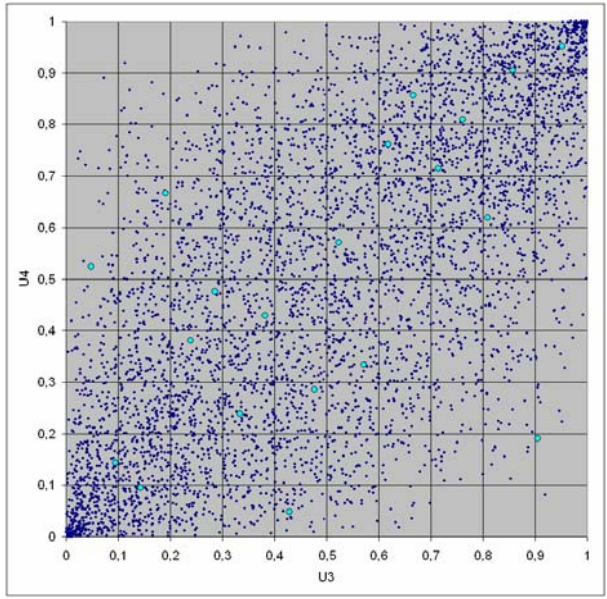 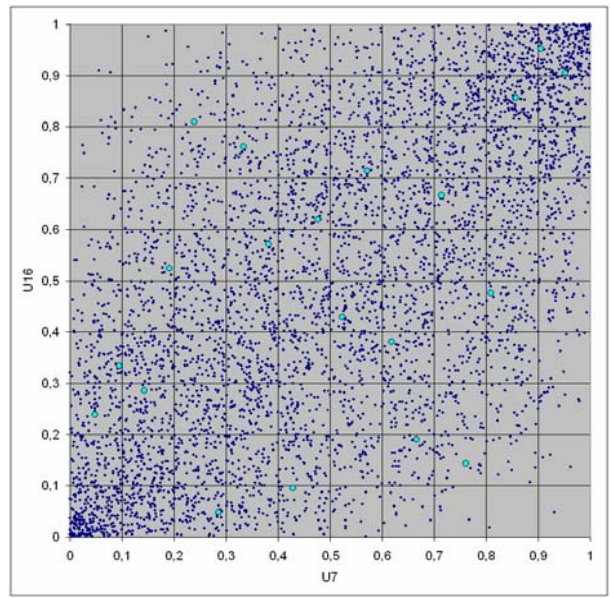

Fig. 34          Fig. 35

Gaussian copula

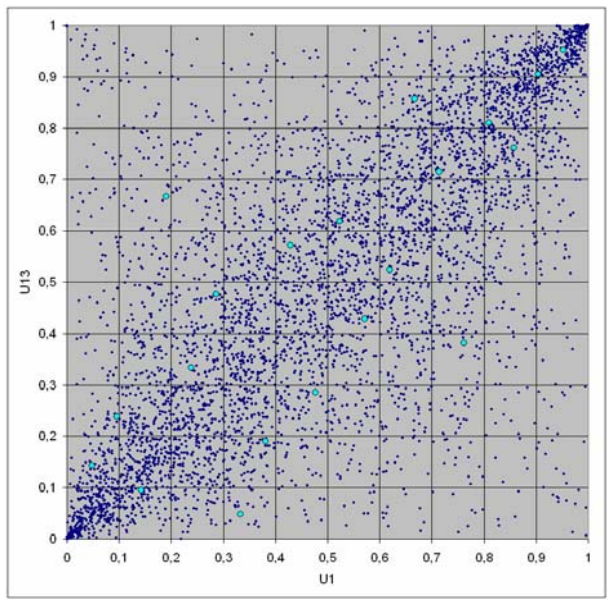 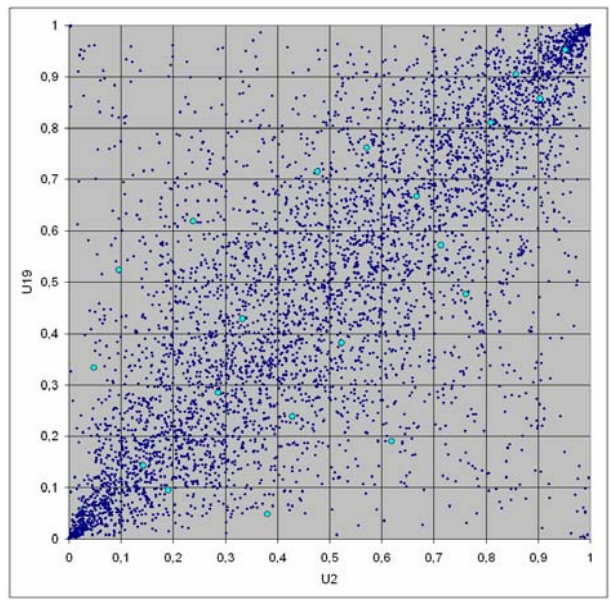

Fig. 36          Fig. 37

t-copula with two degrees of freedom

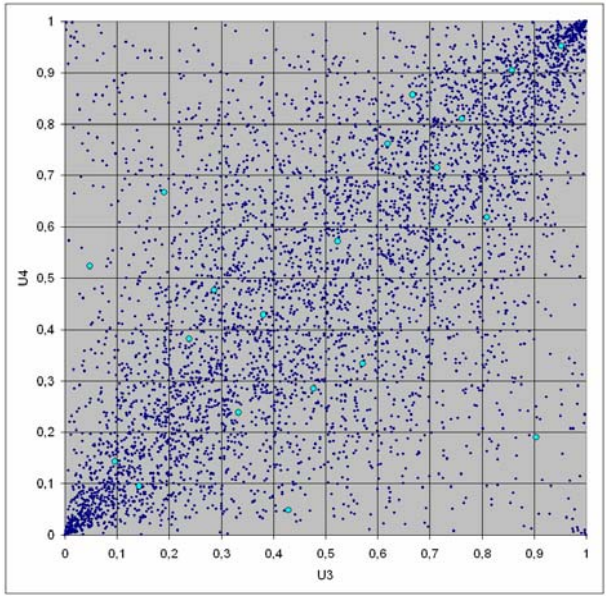
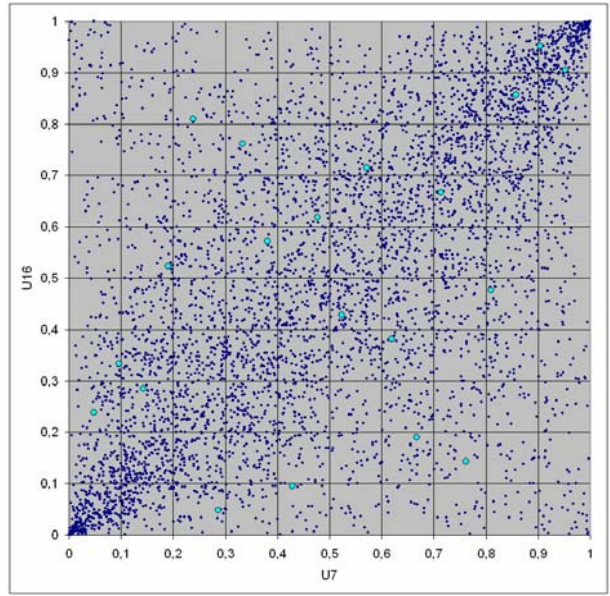

Fig. 38　　　　　　　　　　　　　　　　　Fig. 39

t-copula with two degrees of freedom

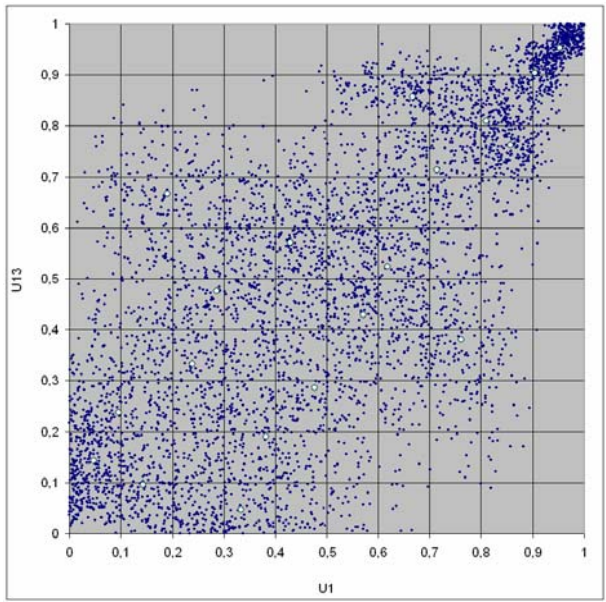
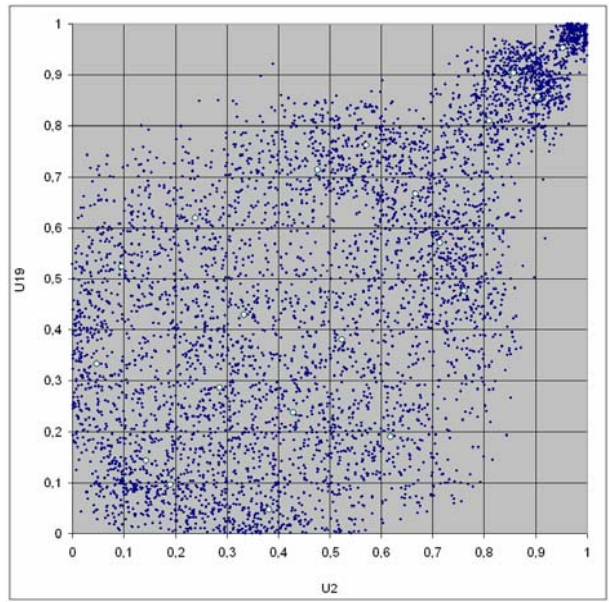

Fig. 40　　　　　　　　　　　　　　　　　Fig. 41

Gamma copula with the rook copula driver

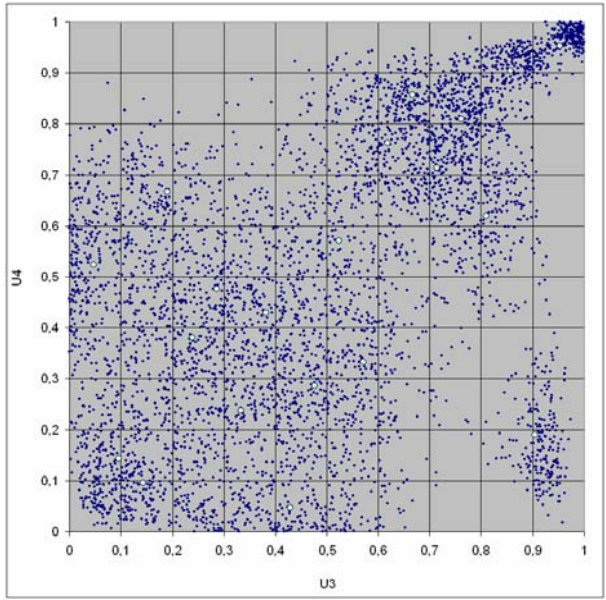
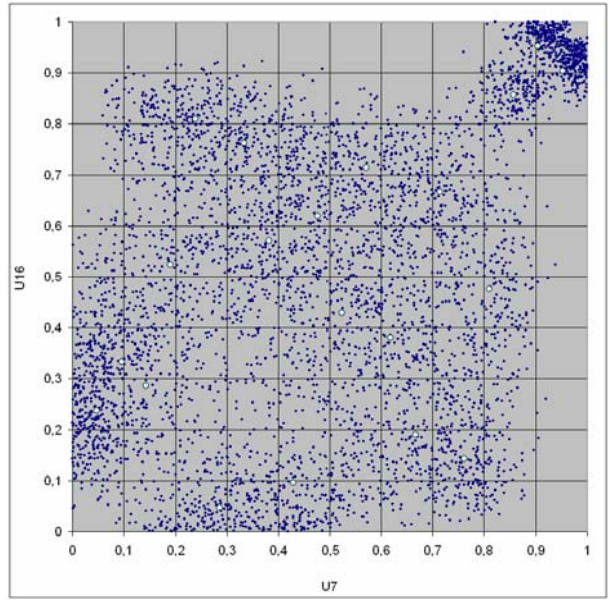

Fig. 42          Fig. 43

Gamma copula with the rook copula driver

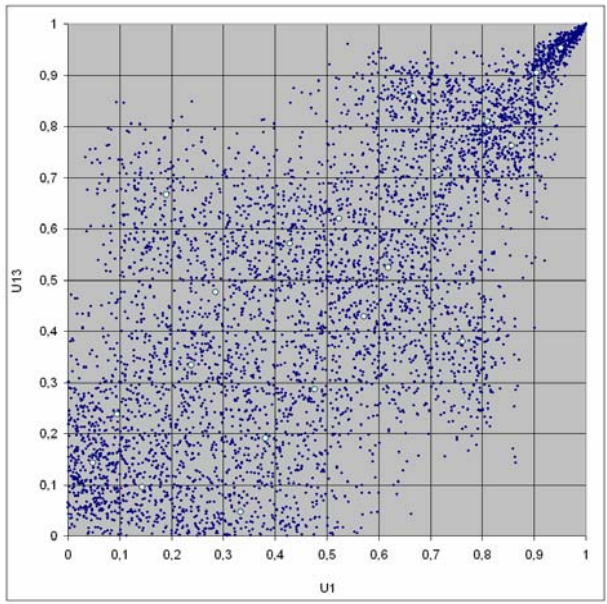
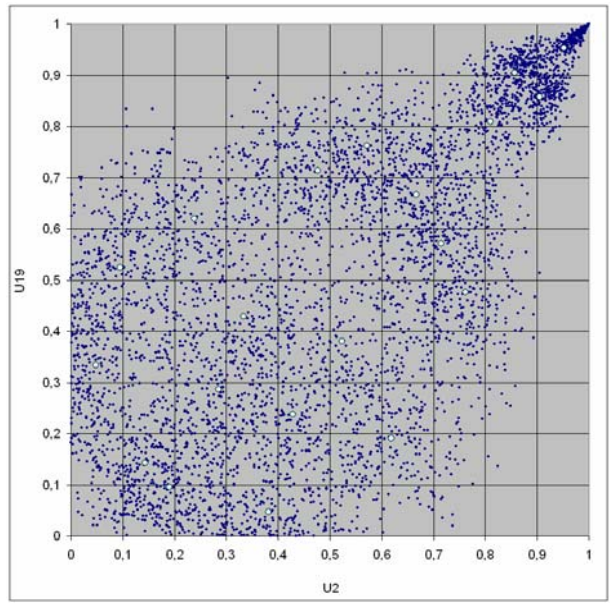

Fig. 44          Fig. 45

Gamma copula with the upper Fréchet copula driver

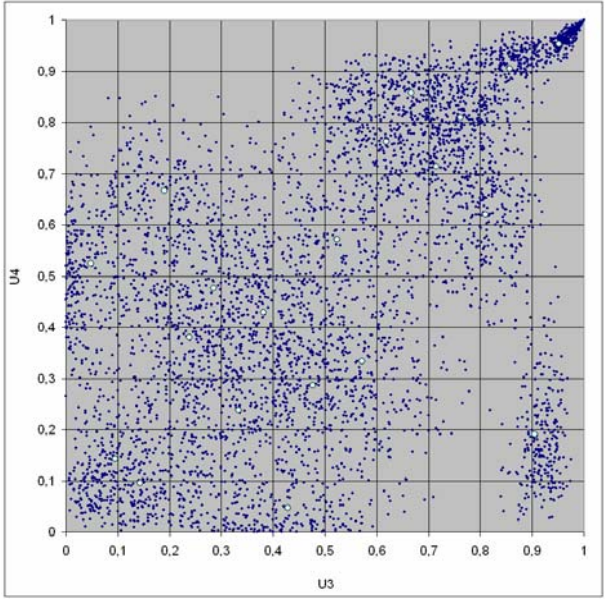 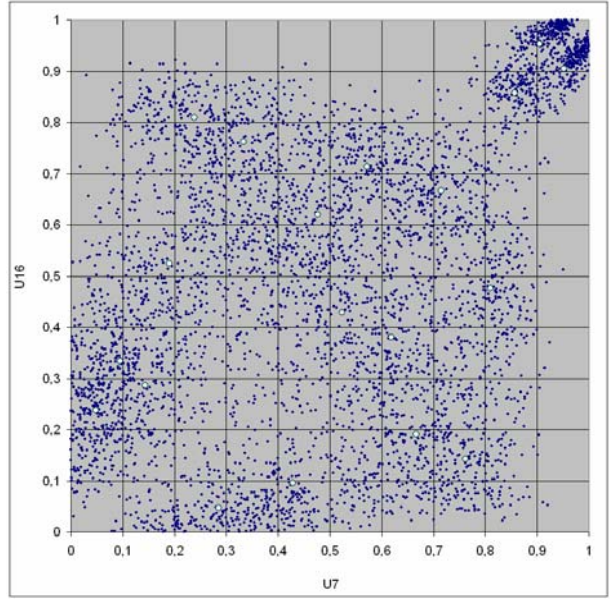

Fig. 46            Fig. 47

Gamma copula with the upper Fréchet copula driver

Obviously, the Gamma copula approach follows the particular asymmetries in the data much better than the Gaussian or *t*-copula constructions. Also, the upper tail dependence seems stronger under the Gamma copula than under the *t*-copula for certain two-dimensional projections.

As an application to risk management, we estimate several values of the risk measure $\mathrm{VaR}_u(S) = Q_S(1-u)$ for different values of *u*, for the aggregate risk $S = \sum_{k=1}^{19} X_k$, where $X_k$ represents the insurance losses in Area *k*. The marginal distributions of the losses were modelled as lognormal, which is in coincidence with the Lilliefors test for the log losses. The analysis is based on 100,000 simulations each. For a direct comparison, note that the Gaussian copula and the Gamma copula with the rook copula driver do not show a tail dependence, whereas the *t*-copula and the Gamma copula with the upper Fréchet copula driver possess a distinct upper tail dependence. We show the results in Tab. 5.

| copula type | Gaussian copula | rook Gamma copula | *t*-copula | UF Gamma copula |
|---|---:|---:|---:|---:|
| $\mathrm{VaR}_{0.1}(S)$ | 828.149 | 1,687.750 | 785.207 | 1,530.999 |
| $\mathrm{VaR}_{0.05}(S)$ | 1,123.028 | 2,097.296 | 1,126.537 | 1,980.437 |
| $\mathrm{VaR}_{0.01}(S)$ | 2,013.425 | 2,865.834 | 2,345.636 | 3,271.872 |
| $\mathrm{VaR}_{0.005}(S)$ | 2,528.785 | 3,283.720 | 3,127.850 | 3,950.194 |

Tab. 5: VaR estimates

Note that the $\mathrm{VaR}_{0.005}(S)$ is the basis for the Solvency Capital Requirement (SCR) under Solvency II, which is more than 50% higher estimated with the UF Gamma copula than with the Gaussian copula, and that even the rook Gamma copula with no tail dependence produces a higher estimate than the *t*-copula which has a tail dependence.

# 6 Final Remarks

We close this paper by the remark that the continuous and discrete finite or infinite partition-of-unity copula approach is absolutely flexible, i.e. it is even possible to choose dimension-wise different distribution families (binomial, negative binomial, Poisson, Gamma, etc.) for the copula estimation and also different copula drivers which can be any reasonable simple patchwork copula based on the observations. Further, it is possible to choose copula drivers which allow for an implementation of tail dependence, even if this feature can in general not be concluded from a finite data set. However, in the light of Solvency II, it might be desirable to compare VaR estimates for aggregate losses with and without a tail dependent copula, and under competing dependence models.

Another advantage is the easy implementation of the simulation algorithm even in ordinary spreadsheet software for arbitrary large dimensions. We have worked in practice with 114-dimensional data sets from the insurance sector without any problems.

# Acknowledgements

We would like to thank the referees for a constructive criticism which led to an improvement in the presentation of the paper.